\documentclass[
    twocolumn,
	prl,
	amssymb,
	preprintnumbers, superscriptaddress,
	nofootinbib]{revtex4-1}

\pdfoutput=1

\usepackage{graphicx}
\usepackage{enumitem}
\usepackage{latexsym}
\usepackage{amsfonts}
\usepackage{amssymb}
\usepackage{xcolor}
\usepackage[export]{adjustbox}
\usepackage{amsmath}
\usepackage{slashed}
\usepackage{dcolumn}
\usepackage{verbatim}
\usepackage{float}
\usepackage{multirow}
\usepackage{xspace}
\usepackage[normalem]{ulem}
\usepackage[
pdfauthor={Jeremy Sakstein}]{hyperref}

\newcommand{\lsim}{\lesssim}

\newcommand{\acro}[1]{\textsc{\MakeLowercase{#1}}} 

\newcommand{\beq}{\begin{equation}}
\newcommand{\eeq}{\end{equation}}
\newcommand{\bea}{\begin{eqnarray}}
\newcommand{\eea}{\end{eqnarray}}

\newcommand{\refjnl}[1]{{\rm#1}}

\def\apj{\refjnl{ApJ}}                 

\def\jcap{\refjnl{J. Cosmology Astropart. Phys.}}
\def\mnras{\refjnl{MNRAS}}             
\newcommand{\msun}{{\rm M}_\odot}

\newcommand{\tenx}[1]{\times 10^{#1}}

\allowdisplaybreaks

\begin{document}

\title{Beyond the Standard Model Explanations of GW190521}
\author{Jeremy Sakstein} \email{sakstein@hawaii.edu}
\affiliation{Department of Physics \& Astronomy, University of Hawai'i, Watanabe Hall, 2505 Correa Road, Honolulu, HI, 96822, USA}
\author{Djuna Croon} \email{dcroon@triumf.ca}
\affiliation{TRIUMF, 4004 Wesbrook Mall, Vancouver, BC V6T 2A3, Canada}
\author{Samuel D.~McDermott}
\email{sammcd00@fnal.gov}
\affiliation{Fermi National Accelerator Laboratory, Batavia, IL USA}
\author{Maria C. Straight}
\email{mstraight21@my.whitworth.edu}
\affiliation{Department of Engineering and Physics, Whitworth University, 300 W. Hawthorne Rd., Spokane, WA 99251}
\author{Eric J. Baxter} 
\email{ebax@hawaii.edu}
\affiliation{Institute  for  Astronomy,  University  of  Hawaii,2680  Woodlawn  Drive,  Honolulu,  HI  96822,  USA}

\preprint{FERMILAB-PUB-20-461-T}

\date{\today}

\begin{abstract}
The LIGO/Virgo collaboration has recently announced the detection of a heavy binary black hole merger, with component masses that cannot be explained by standard stellar structure theory. In this letter we propose several explanations based on models of new physics, including new light particle losses, modified gravity, large extra dimensions, and a small magnetic moment of the neutrino. Each of these affect the physics of the pair-instability differently, leading to novel mechanisms for forming black holes inside the  mass gap. 
\end{abstract}

\maketitle

On 21 May 2019, the Advanced LIGO and Virgo collaboration observed a gravitational wave signal compatible with a binary black hole inspiral of total remnant mass\footnote{The central value and error bars, as will be true throughout, are the median value and the 90\% credible posterior intervals.} $142^{+28}_{-16} {\rm M}_\odot$ \cite{PhysRevLett.125.101102, Abbott_2020}. Remarkably, the component masses, of $m_1 =85^{+21}_{-14}{\rm M}_\odot$ and $m_2=66^{+17}_{-18}{\rm M}_\odot$  respectively, lie firmly within the Black Hole Mass Gap (BHMG) predicted by stellar structure theory. 
While the precise location of the BHMG is subject to various sources of uncertainty --- nuclear, astrophysical, and computational --- the LIGO/Virgo collaboration finds the probability of finding $m_1^{\rm GW190521}$  $<0.1\%$ ($<0.3\%$) if the gap is located at $50{\rm M}_\odot$ ($65{\rm M}_\odot$) \cite{PhysRevLett.125.101102, Abbott_2020}. Recent works \cite{Marchant:2018kun,Farmer:2019jed,Farmer:2020xne,Marchant:2020haw}  found the location of the mass gap to be $\lsim 50 {\rm M}_\odot$ with individual uncertainty estimates no greater than $\pm 4 {\rm M}_\odot$. As such, GW190521 is a potential hint of new physics. 

The BHMG is a manifestation of pair-instability supernovae (PISN): the densities and temperatures in the cores of metal-poor, massive ($\gtrsim 50 {\rm M}_\odot$) population-III stars allow for the production of electron-positron pairs from the plasma. In turn, this reduces the photon pressure, destabilizing the star and causing it to contract. The resulting temperature increase leads to rapid thermonuclear burning of ${}^{16}\textrm{O}$, which releases energy comparable to the star's binding energy. The subsequent explosion can be so violent that the entire star becomes unbound, leaving no black hole remnant. The onset of the PISN defines the lower edge of the BHMG. The pair-instability is quenched in sufficiently heavy objects ($M\gtrsim120{\rm M}_\odot$) because the energy from the contraction is channeled into photodisintegration of heavy elements. The quenching of the PISN defines the upper edge of the BHMG.
 
The presence of physics beyond the Standard Model (BSM) may modify the PISN to produce heavier black holes. Refs. \cite{Croon:2020ehi,Croon:2020oga} pioneered the effect of \emph{light particle emission} during the helium burning stage of the stellar evolution. It was found that an extra dissipation channel at this stage speeds up helium depletion, with the consequence that less oxygen is produced via the ${}^{12}$C$(\alpha,\gamma)^{16}$O reaction. The BSM physics therefore directly implies less violent thermonuclear explosions, resulting in heavier black holes. We will discuss several examples of models of new physics that change the location of the BHMG essentially through this mechanism.

An alternative mechanism is found in models of screened modified gravity. Such theories can effect an environment-dependent modification to the value of the gravitational constant, $G$, which would in turn impact stellar evolution up to and including PISN. As we discuss in more detail below, an increase in $G$ moves both edges of the BHMG to lower masses, while a decrease in $G$ has the opposite effect \cite{MGMG}. GW190521 could therefore result from black holes below the gap that formed in an unscreened environment with decreased $G$, or from black holes above the gap that have formed in an environment with an enhanced $G$.

 \noindent {\bf \emph{GW190521} --}
Following the LIGO/Virgo collaboration, we quote results using the {\tt NRSur7dq4} waveform model \cite{Varma:2019csw}, based on a numerical relativity surrogate model obtained from direct interpolation of numerical relativity models  \cite{Varma:2018mmi} 
(results using other waveform models are compatible for all parameters, with slightly higher median posterior values for all masses \cite{Abbott_2020}).
The total system mass is measured to be $m_1+m_2 = 150^{+29}_{-17}\msun$, its chirp mass $\mathcal M \equiv (m_1 m_2)^{3/5}/(m_1+m_2)^{1/5}$ is $\mathcal M = 64^{+13}_{-8} \msun$ with mass ratio $m_2/m_1 = 0.79^{+0.19}_{-0.29}$, and the initial spins are measured to be $\chi_1 = 0.69^{+0.27}_{-0.62}$ and $\chi_2 = 0.73^{+0.24}_{-0.64}$. The sky localization is $\Delta \Omega = 774{\rm\,deg}^2$, or roughly 2\% of the sky. The source redshift is measured to be $0.82^{+0.28}_{-0.34},$ corresponding to a luminosity distance of $D_{\rm L} = 5.3^{+2.4}_{-2.6}$ Gpc. Using a Jeffreys prior $p(R) \propto R^{-0.5}$ to constrain the rate given a single detection of a rare event, the rate for events similar to GW190521 is $R=0.13^{+0.30}_{-0.11}{\rm\,Gpc^3/yr}$ \cite{Abbott_2020}. The signal is consistent with a circular merger, although the number of orbits in the signal band allows for the possibility of some eccentricity, with unclear implications for the mass extraction \cite{Abbott_2020}.

\noindent{\bf \emph{Standard Model Explanations} --} 
The formation of a black hole within the mass gap is a challenge for standard astrophysical theory. The physics of the pair instability is a simple manifestation of the change in the adiabatic index of a pressure-supported star. Without augmenting the Standard Model (SM) of particle physics with new phenomena, there are two pathways for producing black holes in the mass gap: quenching stellar winds or pulsations {\it before} the black hole is formed, or adding mass {\it after} the black hole is formed. Using an old version of the numerical code {\tt{MESA}} and the assumption that the entire  He+C+O stellar core after pulsations becomes a black hole -- in reality, only bound material contributes -- Ref.~\cite{Leung:2019fgj} reports a black hole mass slightly in excess of the value found in Ref.~\cite{Farmer:2019jed}. Likewise, Ref.~\cite{Mapelli:2019ipt} reports a remnant mass of $56\msun$ for a single, non-rotating progenitor with an intact hydrogen shell using the model of Ref.~\cite{Fryer:2011cx} to diagnose a failed explosion. However, using the most up-to-date version of {\tt{MESA}}, uncertain and stochastic effects like convection \cite{2020MNRAS.493.4333R}, wind-loss efficiency, mixing, electroweak physics, and metallicity \cite{Farmer:2019jed} were all found to affect $M_{\rm BHMG}^{\rm(SM)} \lesssim 50 \msun$, the location of the lower boundary of the mass gap in the SM, by $\lesssim 3\msun$. 
One seeming exception to this rule is the uncertainty derived from the $^{12}$C$(\alpha,\gamma)^{16}$O rate, variation of which can substantially impact $M_{\rm BHMG}^{\rm(SM)}$ because of its importance in determining the amount of combustible ${}^{16}$O present during pulsations \cite{Farmer:2020xne}. Nevertheless, the most recent compilation of experimental and theoretical work on this rate \cite{deBoer:2017ldl} leads to a propagated $1\sigma$ uncertainty on $M_{\rm BHMG}^{\rm(SM)}$ of only $+0(-4)\msun$ \cite{Farmer:2020xne}. Thus, this rate is not expected to lead to substantially larger uncertainty compared to other SM effects \cite{nuclearconvo}, and the direct creation of a black hole, especially one as massive as the primary object in GW190521, seems unlikely.

The formation of black holes in the mass gap {\it after} the formation of smaller initial black holes is possible through two primary mechanisms. One possibility is hierarchical mergers of two sub-mass-gap black holes. As discussed in detail in \cite{Abbott_2020}, the likelihood of hierarchical mergers depends strongly on the natal environment and the details of the initial black hole formation, but a larger number of high-precision events will enable statistical analysis of these effects. Alternatively, super-Eddington accretion (from common-envelope evolution or stable mass transfer) may ``pollute'' the mass gap \cite{Roupas:2018cvb,Roupas:2019dgx}, but a recent study in the context of a population synthesis model found no pairs with combined mass in excess of $100\msun$ \cite{2020ApJ...897..100V}. These mechanisms for mass increase after black hole birth, though beset by uncertainties, will be better understood with dedicated modeling and additional observational constraints from future LIGO/Virgo runs.

\begin{figure*}[t]
    \centering
    {\includegraphics[width=\textwidth]{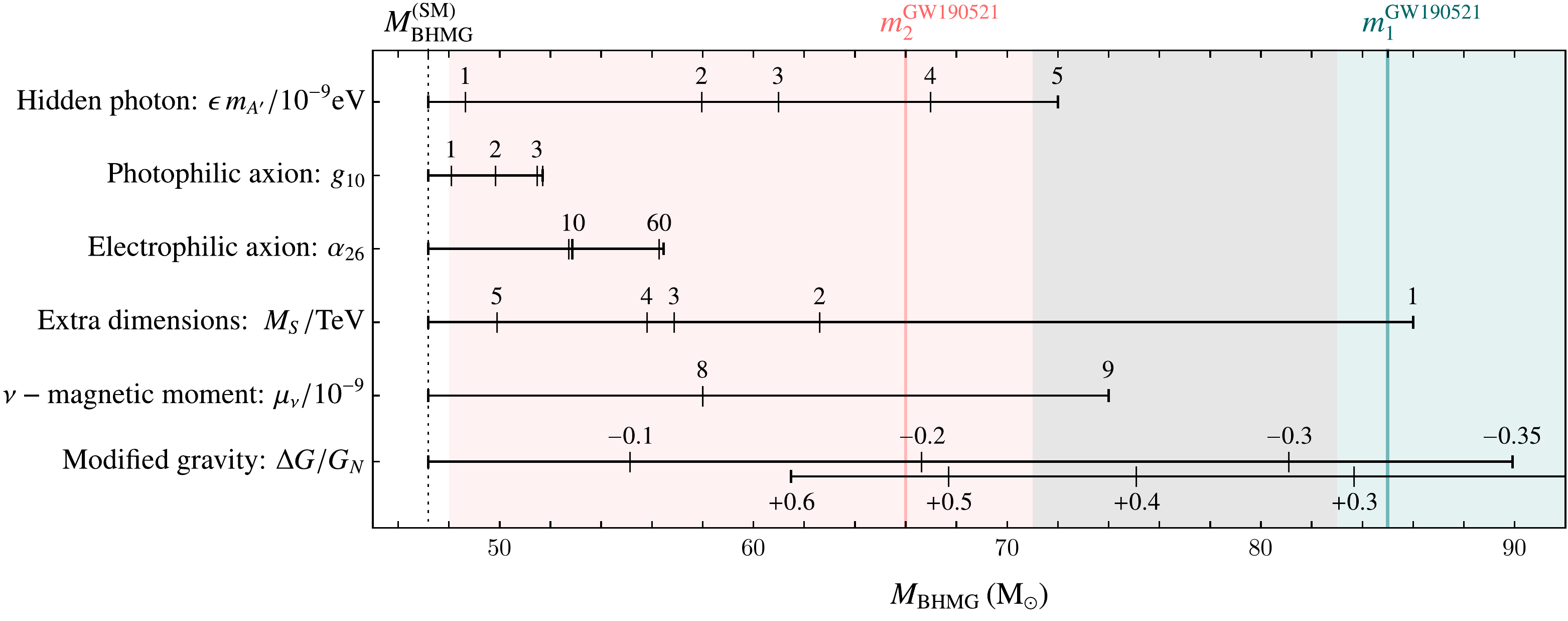}}
    \caption{Beyond the Standard Model explanations of GW190521 --- the lower edge of the Black Hole Mass Gap $M_{\rm BHMG}$ in various theories of new physics, as well as the upper edge in the case of modified gravity. In cyan, the GW190521 primary black hole mass $m_1 =85^{+21}_{-14}{\rm M}_\odot$ and in pink, the secondary mass $m_2=66^{+17}_{-18}{\rm M}_\odot$. The dashed line gives the Standard Model prediction for the lower edge of the Black Hole Mass Gap, $M_{\rm BHMG}^{\rm (SM)}$.}
    \label{fig:summary}
\end{figure*}

\vspace{1mm}
\noindent{\bf \emph{Beyond the Standard Model Explanations} --}
While uncertainties, predominantly astrophysical in origin, preclude unambiguous interpretation of GW190521, we demonstrate here how BSM physics can bridge the gap and produce black holes with masses consistent with those observed in GW190521. Our focus here is on new light particles, large extra dimensions, a large magnetic moment of active neutrinos, and modified gravity, but other BSM models may affect the PISN through the same essential mechanisms. Our quantitative results derive from numerical simulations using the stellar structure code {\tt MESA} version 12778 (the most up-to-date version) \cite{Paxton2011-MESA,Paxton:2015jva,Paxton2018-MESA,Roupas:2019dgx}. Our numerical prescription is given in \cite{Croon:2020ehi,Croon:2020oga,MGMG} and mirrors recent studies of the BHMG using the same code \cite{Marchant:2018kun,Farmer:2019jed}. We summarize our results in Fig.~\ref{fig:summary}.

\noindent {\bf New particles} are abundant in BSM theories and may potentially be related to dark matter physics. Light degrees of freedom which couple to the material in the star may lead to energy loss, as neutrinos do in the SM.
The effect that this extra dissipation has on $M_{\rm BHMG}$ depends on the production mechanism.
Refs.~\cite{Croon:2020ehi,Croon:2020oga} studied the effects of hidden photons, primarily produced via plasma resonance, photophilic axions, primarily produced via Primakov emission, and electrophilic axions,\footnote{Of particular interest in light of an excess at XENON1T \cite{Aprile:2020tmw,Croon:2020ehi}.} produced primarily in the semi-Compton process $e +\gamma \to e+ \gamma +a$. The different temperature scaling of these effects ($\propto \epsilon^2 m_{A'}^2 \,T$, $\propto g_{10}^2\, T^4$, $\propto \alpha_{26}\, T^6$ for hidden photons, photophilic axions, and electrophilic axions respectively) implies a different scaling of $M_{\rm BHMG}$ with BSM coupling. 

Dissipation due to light particle production in stellar cores has been studied in other astrophysical scenarios, resulting in stringent constraints. In particular, the 
inferred cooling rates of white dwarfs and red giants in globular clusters \cite{DiLuzio:2020jjp} provide constraints in tension with the electrophilic axion scenario presented here; observations of the sun with the CAST experiment \cite{Anastassopoulos:2017ftl} and from global structure fits \cite{2015JCAP...10..015V} constrain the photophilic axion to have a coupling $g_{10} \leq 4$ or stronger, depending on the axion mass; and requiring that dark photons not overwhelm the luminosity of Boron-8 neutrinos from the Sun places a limit $\epsilon m_{A'} \lesssim 1\tenx{-11}$ eV \cite{An:2013yfc,Redondo:2013lna}. However, underexplored parameter degeneracies such as the population metallicity \cite{Gao:2020wer} or the age of the stars may reduce the tension in distant stellar systems, and a new particle with a mass kinematically accessible to helium-burning but not hydrogen-burning reactions could evade Solar bounds. We consider the electrophilic axion to be the most promising candidate for further study, but analyses including propagation of all uncertainties of these stellar environments and with the full ranges of particle properties is needed to definitively understand whether or not the particles we suggest as drivers of large $M_{\rm BHMG}$ are phenomenologically viable.

\noindent {\bf Large extra dimensions} may play a key role in understanding cosmological mysteries such as the nature of dark energy \cite{Copeland:2006wr} and the cosmological constant problem \cite{Burgess:2013ara,Padilla:2015aaa,Khoury:2018vdv}. In these scenarios, the SM is confined to a four-dimensional sub-manifold embedded in a higher dimensional bulk space-time governed by higher-dimensional gravity. This allows for a novel loss mechanism in the interiors of stars whereby SM particles can produce Kaluza-Klein (KK) modes of the graviton. We have implemented the losses due to KK modes resulting from photon-photon annihilation, gravi-Compton-Primakoff scattering, and gravi-bremsstrahlung processes into {\tt{MESA}} using the rates given in \cite{Mori:2020niw} for $n$ additional compactified spatial dimensions. The most important mechanism is photon-photon annihilation, which scales as $T^{n+7}$.

The case $n=1$ is strongly constrained by solar system tests and for $n\geq3$ the constraints from collider experiments are expected to be substantially stronger than those available from stellar evolution \cite{Mori:2020niw}, so we focus on $n=2$. The effects of losses are similar to those resulting from light particle emission: the ratio of $^{12}$C to $^{16}$O after core helium burning is enhanced, quenching the PISN and forming heavier black holes. With the constraint $R_{\rm LED} \leq 30 \mu$m \cite{Tan:2020vpf,Lee:2020zjt,Mori:2020niw}, the LED scenario is capable of accounting for the GW190521 progenitor masses if the mass scale associated with the extra dimensions satisfies $M_s<1-2$ TeV for $n=2$. 
For $n=2$, CMS constrains $M_s \gtrsim 3.9$ TeV \cite{Sirunyan:2017jix}, which is again in some tension with the value needed to produce a black hole as massive as the primary object in GW190521, though this constraint depends on the compactification geometry and details of the collider signal process, and the viability of compactifications with mass scale $\lesssim 2$ TeV calls for a thorough investigation.

\noindent {\bf The magnetic dipole moment of the neutrino} is very small in the SM, such that for $\mathcal L \supset \frac{\mu_\nu}2 \bar \nu \sigma_{\alpha \beta} \nu F^{\alpha \beta}$ the SM expectation is $\mu_\nu \sim \mathcal O(10^{-19})\mu_B$ \cite{Balantekin:2013sda}, where $\mu_B=e/2m_e$ is the Bohr magneton. However, this operator can be large in BSM scenarios, which predict values as large as $\mu_\nu \sim \mathcal O(10^{-10})\mu_B$. The non-zero magnetic moment implies a coupling of the photon to neutrinos, leading to additional neutrino production channels inside plasmas. As neutrinos are light enough to escape the star's gravitational field, the enhanced emissivity in the stellar interior results in similar effects to those of new light particle emission.

We follow Refs.~\cite{Barger:1999jf,Heger:2008er} to implement the additional neutrino losses: plasmon decay ($\gamma \to \nu \bar\nu$) and pair production via electron-positron annihilation ($e^+e^- \to \nu \bar\nu$). As the latter process is strongly suppressed until very high temperatures and quenched for a degenerate plasma, the plasmon decay is expected to dominate during the evolution of interest.
We performed a full investigation into the resulting suppression of the PISN with {\tt{MESA}} and found that black holes with masses compatible with the GW190521 progenitors can be formed only if the neutrino's magnetic moment exceeds $\mu_\nu \gtrsim10^{-9}\mu_B$. Such large magnetic moments are necessary as the enhanced emission dominates at high density and low temperatures \cite{Heger:2008er}, away from typical population-III stellar tracks.
Observations of red giant stars in globular clusters place the leading constraints on such models, as strong as $\mu_\nu< 2.2 \times 10^{-12}\mu_B$ \cite{Raffelt:1999tx}, and low-energy laboratory experiments constrain values of $\mu_\nu$ of this magnitude as well \cite{Giunti:2014ixa}. As such, we do not consider the neutrino magnetic moment to be a viable explanation of GW190521.

\noindent {\bf Modified gravity} theories are able to simultaneously drive the acceleration of the cosmic expansion (dark energy) while remaining compatible with solar system tests of gravity \cite{Joyce:2014kja,Burrage:2017qrf,Sakstein2018-testScreened,Baker:2019gxo}. They are also leading candidates for explaining the Hubble tension \cite{Sakstein2019-Screened,Desmond2019-resolveHT,Desmond2020-Screened}. Their success is due to screening mechanisms, which suppress fifth forces in high density environments. Low density objects such as dwarf galaxies in cosmic voids may be unscreened, leading to large (100\% or more) variations in the local strength of gravity. If $G_{\rm N}$ is the value of Newton's constant measured in the solar system, then objects in unscreened regions experience an effective value of the gravitational constant parameterized by $G=G_{\rm N}(1+\Delta G/G_{\rm N})$, with $\Delta G/G_{\rm N}$  environment-dependent. Screened environments like the Milky Way therefore have $\Delta G/G_{\rm N}=0$.

In a forthcoming publication, three of us will present the results of a detailed investigation \cite{MGMG} into the effects of such theories on the structure, evolution, and fate of population-III stars. We find that increasing the local strength of gravity has the effect of raising the core temperature at fixed density due to the necessity of a larger pressure gradient to maintain hydrostatic equilibrium. This increases the available energy for electron-positron pair production, exacerbating the pair-instability. The stars then undergo more violent pulsations encounter the instability at lower mass, which prevents the formation of heavy black holes. Similarly, the same effect causes the PISN to quench in lower mass objects, leading to the formation of intermediate mass ($\sim 60$--$90{\rm M}_\odot$) black holes that would otherwise have gone PISN. The ultimate result is a decrease in the location of both the upper and lower edges of the BHMG. The converse is true for theories where the strength of gravity is reduced. 
 
For values of $\Delta G/G_{\rm N}>0.25$, upper edge black holes have masses compatible with the primaries observed in GW190521. Interestingly, this value (specifically $\Delta G/G_{\rm N}=1/3$) is typical of chameleon and $f(R)$ theories. It is thus possible that the progenitors of GW190521 formed from heavier population-III stars in an unscreened galaxy, while other LIGO/Virgo black holes formed in screened galaxies, consistent with the general relativity (GR) prediction of  $M_{\rm BHMG}^{\rm (SM)}$. Similarly, for values of $\Delta G/G_{\rm N}\lsim-0.30$, the lower edge of the BHMG lies at $M_{\rm BHMG} \sim 85{\rm M}_\odot$. Note that the dynamics of the black hole merger is identical to the corresponding GR process. Screened modified gravity theories introduce new light degrees of freedom (typically scalars) and, as such, are subject to no-hair theorems \cite{Hui:2012qt,Sotiriou:2013qea}. For this reason, the black hole solutions and dynamics are identical to the predictions of GR: only the progenitor stars feel the fifth force. 

Theories in which the strength of gravity weakens in unscreened environments are less common, with beyond Horndeski models as the leading candidates \cite{Koyama:2015oma,Saito:2015fza}. The parameters of these theories needed to account for $m_1$ and $m_2$ inferred from GW190521 are nominally excluded by other stellar structure probes \cite{Koyama:2015oma,Saito:2015fza,Sakstein:2015zoa,Sakstein:2015aac,Sakstein:2016lyj,Babichev:2016jom,Sakstein:2016oel,Sakstein2018-testScreened,Saltas:2019ius}, however the parameter $\Upsilon_1$ that controls the modifications is redshift-dependent. Since the current constraints derive from low-$z$ probes, it is in principle possible to construct theories that imply strong deviations at $5.3$ Gpc while being consistent with local constraints.

\vspace{1mm} \noindent
{\bf \emph{Discussion} --}
Simulating the six models of new physics discussed in the previous section, we derive the lower edge of the black hole mass gap, $M_{\rm BHMG}$. The results of these calculations are given in Fig.~\ref{fig:summary}. This figure also shows the best fit values and $90\%$ confidence intervals of the GW190521 component masses. 

As seen in Fig.~\ref{fig:summary}, the tension between the component masses of GW190521 and the SM prediction of $M_{\rm BHMG}^{\rm (SM)}$ can be (partially) mitigated in models of new physics. In particular hidden photon emission, large extra dimensions, and modified gravity may explain the result at the $2\sigma$-level, assuming that the SM result is $M_{\rm BHMG}^{\rm (SM)}=47.2{\rm M}_\odot$. We note that SM uncertainties are largely independent and complementary to the BSM effects presented here, such that a combination of different effects may ultimately bridge the gap. 

The exceptional event GW190521 provides a tantalizing hint of new physics at work in pair-instability supernovae. The full catalogue of O3 events may be used to shed further light on the black hole mass gap. 
While the location and shape of the black hole population remains uncertain, future gravitational wave observations may reveal correlations between BH masses, spins, merger rate, and redshift, and disentangle the various theoretical uncertainties from the new phenomenology proposed here.
As the LIGO/Virgo collaborations continue their thrilling exploration of new, unexpected corners of the black hole landscape, we are excited to anticipate the implications of these ``black hole colliders'' for particle physics beyond the Standard Model.

\noindent {\bf Acknowledgements --}  We thank Toby Opferkuch for useful discussions, we are grateful to Robert Farmer, Pablo Marchant, and the wider {\tt{MESA}} community for answering our many questions, and we are especially thankful to the LIGO-Virgo-Kagra collaborations for their hard work making this event public; we commend the gravitational wave community for inspiring us all.
T\acro{RIUMF} receives federal funding via a contribution agreement with the National Research Council Canada.
Fermilab is operated by Fermi Research Alliance, LLC under Contract No. De-AC02-07CH11359 with the United States Department of Energy. 
MCS acknowledges support from the Research Experience for Undergraduate program at the Institute for Astronomy, University of Hawai'i-Manoa funded through NSF grant 6104374.

\bibliography{refs}

\begin{thebibliography}{60}%
\makeatletter
\providecommand \@ifxundefined [1]{%
 \@ifx{#1\undefined}
}%
\providecommand \@ifnum [1]{%
 \ifnum #1\expandafter \@firstoftwo
 \else \expandafter \@secondoftwo
 \fi
}%
\providecommand \@ifx [1]{%
 \ifx #1\expandafter \@firstoftwo
 \else \expandafter \@secondoftwo
 \fi
}%
\providecommand \natexlab [1]{#1}%
\providecommand \enquote  [1]{``#1''}%
\providecommand \bibnamefont  [1]{#1}%
\providecommand \bibfnamefont [1]{#1}%
\providecommand \citenamefont [1]{#1}%
\providecommand \href@noop [0]{\@secondoftwo}%
\providecommand \href [0]{\begingroup \@sanitize@url \@href}%
\providecommand \@href[1]{\@@startlink{#1}\@@href}%
\providecommand \@@href[1]{\endgroup#1\@@endlink}%
\providecommand \@sanitize@url [0]{\catcode `\\12\catcode `\$12\catcode
  `\&12\catcode `\#12\catcode `\^12\catcode `\_12\catcode `\%12\relax}%
\providecommand \@@startlink[1]{}%
\providecommand \@@endlink[0]{}%
\providecommand \url  [0]{\begingroup\@sanitize@url \@url }%
\providecommand \@url [1]{\endgroup\@href {#1}{\urlprefix }}%
\providecommand \urlprefix  [0]{URL }%
\providecommand \Eprint [0]{\href }%
\providecommand \doibase [0]{http://dx.doi.org/}%
\providecommand \selectlanguage [0]{\@gobble}%
\providecommand \bibinfo  [0]{\@secondoftwo}%
\providecommand \bibfield  [0]{\@secondoftwo}%
\providecommand \translation [1]{[#1]}%
\providecommand \BibitemOpen [0]{}%
\providecommand \bibitemStop [0]{}%
\providecommand \bibitemNoStop [0]{.\EOS\space}%
\providecommand \EOS [0]{\spacefactor3000\relax}%
\providecommand \BibitemShut  [1]{\csname bibitem#1\endcsname}%
\let\auto@bib@innerbib\@empty
\bibitem [{\citenamefont {Abbott}\ \emph
  {et~al.}(2020{\natexlab{a}})\citenamefont {Abbott} \emph
  {et~al.}}]{PhysRevLett.125.101102}%
  \BibitemOpen
  \bibfield  {author} {\bibinfo {author} {\bibfnamefont {R.}~\bibnamefont
  {Abbott}} \emph {et~al.} (\bibinfo {collaboration} {LIGO Scientific
  Collaboration and Virgo Collaboration}),\ }\href {\doibase
  10.1103/PhysRevLett.125.101102} {\bibfield  {journal} {\bibinfo  {journal}
  {Phys. Rev. Lett.}\ }\textbf {\bibinfo {volume} {125}},\ \bibinfo {pages}
  {101102} (\bibinfo {year} {2020}{\natexlab{a}})}\BibitemShut {NoStop}%
\bibitem [{\citenamefont {Abbott}\ \emph
  {et~al.}(2020{\natexlab{b}})\citenamefont {Abbott} \emph
  {et~al.}}]{Abbott_2020}%
  \BibitemOpen
  \bibfield  {author} {\bibinfo {author} {\bibfnamefont {R.}~\bibnamefont
  {Abbott}} \emph {et~al.},\ }\href {\doibase 10.3847/2041-8213/aba493}
  {\bibfield  {journal} {\bibinfo  {journal} {The Astrophysical Journal}\
  }\textbf {\bibinfo {volume} {900}},\ \bibinfo {pages} {L13} (\bibinfo {year}
  {2020}{\natexlab{b}})}\BibitemShut {NoStop}%
\bibitem [{\citenamefont {Marchant}\ \emph {et~al.}(2018)\citenamefont
  {Marchant}, \citenamefont {Renzo}, \citenamefont {Farmer}, \citenamefont
  {Pappas}, \citenamefont {Taam}, \citenamefont {de~Mink},\ and\ \citenamefont
  {Kalogera}}]{Marchant:2018kun}%
  \BibitemOpen
  \bibfield  {author} {\bibinfo {author} {\bibfnamefont {P.}~\bibnamefont
  {Marchant}}, \bibinfo {author} {\bibfnamefont {M.}~\bibnamefont {Renzo}},
  \bibinfo {author} {\bibfnamefont {R.}~\bibnamefont {Farmer}}, \bibinfo
  {author} {\bibfnamefont {K.~M.}\ \bibnamefont {Pappas}}, \bibinfo {author}
  {\bibfnamefont {R.~E.}\ \bibnamefont {Taam}}, \bibinfo {author}
  {\bibfnamefont {S.}~\bibnamefont {de~Mink}}, \ and\ \bibinfo {author}
  {\bibfnamefont {V.}~\bibnamefont {Kalogera}},\ }\href {\doibase
  10.3847/1538-4357/ab3426} {\  (\bibinfo {year} {2018}),\
  10.3847/1538-4357/ab3426},\ \Eprint {http://arxiv.org/abs/1810.13412}
  {arXiv:1810.13412 [astro-ph.HE]} \BibitemShut {NoStop}%
\bibitem [{\citenamefont {Farmer}\ \emph {et~al.}(2019)\citenamefont {Farmer},
  \citenamefont {Renzo}, \citenamefont {de~Mink}, \citenamefont {Marchant},\
  and\ \citenamefont {Justham}}]{Farmer:2019jed}%
  \BibitemOpen
  \bibfield  {author} {\bibinfo {author} {\bibfnamefont {R.}~\bibnamefont
  {Farmer}}, \bibinfo {author} {\bibfnamefont {M.}~\bibnamefont {Renzo}},
  \bibinfo {author} {\bibfnamefont {S.}~\bibnamefont {de~Mink}}, \bibinfo
  {author} {\bibfnamefont {P.}~\bibnamefont {Marchant}}, \ and\ \bibinfo
  {author} {\bibfnamefont {S.}~\bibnamefont {Justham}},\ }\href {\doibase
  10.3847/1538-4357/ab518b} {\  (\bibinfo {year} {2019}),\
  10.3847/1538-4357/ab518b},\ \Eprint {http://arxiv.org/abs/1910.12874}
  {arXiv:1910.12874 [astro-ph.SR]} \BibitemShut {NoStop}%
\bibitem [{\citenamefont {Farmer}\ \emph {et~al.}(2020)\citenamefont {Farmer},
  \citenamefont {Renzo}, \citenamefont {de~Mink}, \citenamefont {Fishbach},\
  and\ \citenamefont {Justham}}]{Farmer:2020xne}%
  \BibitemOpen
  \bibfield  {author} {\bibinfo {author} {\bibfnamefont {R.}~\bibnamefont
  {Farmer}}, \bibinfo {author} {\bibfnamefont {M.}~\bibnamefont {Renzo}},
  \bibinfo {author} {\bibfnamefont {S.}~\bibnamefont {de~Mink}}, \bibinfo
  {author} {\bibfnamefont {M.}~\bibnamefont {Fishbach}}, \ and\ \bibinfo
  {author} {\bibfnamefont {S.}~\bibnamefont {Justham}},\ }\href@noop {} {\
  (\bibinfo {year} {2020})},\ \Eprint {http://arxiv.org/abs/2006.06678}
  {arXiv:2006.06678 [astro-ph.HE]} \BibitemShut {NoStop}%
\bibitem [{\citenamefont {Marchant}\ and\ \citenamefont
  {Moriya}(2020)}]{Marchant:2020haw}%
  \BibitemOpen
  \bibfield  {author} {\bibinfo {author} {\bibfnamefont {P.}~\bibnamefont
  {Marchant}}\ and\ \bibinfo {author} {\bibfnamefont {T.}~\bibnamefont
  {Moriya}},\ }\href@noop {} {\  (\bibinfo {year} {2020})},\ \Eprint
  {http://arxiv.org/abs/2007.06220} {arXiv:2007.06220 [astro-ph.HE]}
  \BibitemShut {NoStop}%
\bibitem [{\citenamefont {Croon}\ \emph
  {et~al.}(2020{\natexlab{a}})\citenamefont {Croon}, \citenamefont
  {McDermott},\ and\ \citenamefont {Sakstein}}]{Croon:2020ehi}%
  \BibitemOpen
  \bibfield  {author} {\bibinfo {author} {\bibfnamefont {D.}~\bibnamefont
  {Croon}}, \bibinfo {author} {\bibfnamefont {S.~D.}\ \bibnamefont
  {McDermott}}, \ and\ \bibinfo {author} {\bibfnamefont {J.}~\bibnamefont
  {Sakstein}},\ }\href@noop {} {\  (\bibinfo {year} {2020}{\natexlab{a}})},\
  \Eprint {http://arxiv.org/abs/2007.00650} {arXiv:2007.00650 [hep-ph]}
  \BibitemShut {NoStop}%
\bibitem [{\citenamefont {Croon}\ \emph
  {et~al.}(2020{\natexlab{b}})\citenamefont {Croon}, \citenamefont
  {McDermott},\ and\ \citenamefont {Sakstein}}]{Croon:2020oga}%
  \BibitemOpen
  \bibfield  {author} {\bibinfo {author} {\bibfnamefont {D.}~\bibnamefont
  {Croon}}, \bibinfo {author} {\bibfnamefont {S.~D.}\ \bibnamefont
  {McDermott}}, \ and\ \bibinfo {author} {\bibfnamefont {J.}~\bibnamefont
  {Sakstein}},\ }\href@noop {} {\  (\bibinfo {year} {2020}{\natexlab{b}})},\
  \Eprint {http://arxiv.org/abs/2007.07889} {arXiv:2007.07889 [gr-qc]}
  \BibitemShut {NoStop}%
\bibitem [{\citenamefont {Straight}\ \emph {et~al.}(prep)\citenamefont
  {Straight}, \citenamefont {Sakstein},\ and\ \citenamefont {Baxter}}]{MGMG}%
  \BibitemOpen
  \bibfield  {author} {\bibinfo {author} {\bibfnamefont {M.}~\bibnamefont
  {Straight}}, \bibinfo {author} {\bibfnamefont {J.}~\bibnamefont {Sakstein}},
  \ and\ \bibinfo {author} {\bibfnamefont {E.}~\bibnamefont {Baxter}},\
  }\href@noop {} {\  (\bibinfo {year} {in prep.})}\BibitemShut {NoStop}%
\bibitem [{\citenamefont {Varma}\ \emph
  {et~al.}(2019{\natexlab{a}})\citenamefont {Varma}, \citenamefont {Field},
  \citenamefont {Scheel}, \citenamefont {Blackman}, \citenamefont {Gerosa},
  \citenamefont {Stein}, \citenamefont {Kidder},\ and\ \citenamefont
  {Pfeiffer}}]{Varma:2019csw}%
  \BibitemOpen
  \bibfield  {author} {\bibinfo {author} {\bibfnamefont {V.}~\bibnamefont
  {Varma}}, \bibinfo {author} {\bibfnamefont {S.~E.}\ \bibnamefont {Field}},
  \bibinfo {author} {\bibfnamefont {M.~A.}\ \bibnamefont {Scheel}}, \bibinfo
  {author} {\bibfnamefont {J.}~\bibnamefont {Blackman}}, \bibinfo {author}
  {\bibfnamefont {D.}~\bibnamefont {Gerosa}}, \bibinfo {author} {\bibfnamefont
  {L.~C.}\ \bibnamefont {Stein}}, \bibinfo {author} {\bibfnamefont {L.~E.}\
  \bibnamefont {Kidder}}, \ and\ \bibinfo {author} {\bibfnamefont {H.~P.}\
  \bibnamefont {Pfeiffer}},\ }\href {\doibase 10.1103/PhysRevResearch.1.033015}
  {\bibfield  {journal} {\bibinfo  {journal} {Phys. Rev. Research.}\ }\textbf
  {\bibinfo {volume} {1}},\ \bibinfo {pages} {033015} (\bibinfo {year}
  {2019}{\natexlab{a}})},\ \Eprint {http://arxiv.org/abs/1905.09300}
  {arXiv:1905.09300 [gr-qc]} \BibitemShut {NoStop}%
\bibitem [{\citenamefont {Varma}\ \emph
  {et~al.}(2019{\natexlab{b}})\citenamefont {Varma}, \citenamefont {Field},
  \citenamefont {Scheel}, \citenamefont {Blackman}, \citenamefont {Kidder},\
  and\ \citenamefont {Pfeiffer}}]{Varma:2018mmi}%
  \BibitemOpen
  \bibfield  {author} {\bibinfo {author} {\bibfnamefont {V.}~\bibnamefont
  {Varma}}, \bibinfo {author} {\bibfnamefont {S.~E.}\ \bibnamefont {Field}},
  \bibinfo {author} {\bibfnamefont {M.~A.}\ \bibnamefont {Scheel}}, \bibinfo
  {author} {\bibfnamefont {J.}~\bibnamefont {Blackman}}, \bibinfo {author}
  {\bibfnamefont {L.~E.}\ \bibnamefont {Kidder}}, \ and\ \bibinfo {author}
  {\bibfnamefont {H.~P.}\ \bibnamefont {Pfeiffer}},\ }\href {\doibase
  10.1103/PhysRevD.99.064045} {\bibfield  {journal} {\bibinfo  {journal} {Phys.
  Rev. D}\ }\textbf {\bibinfo {volume} {99}},\ \bibinfo {pages} {064045}
  (\bibinfo {year} {2019}{\natexlab{b}})},\ \Eprint
  {http://arxiv.org/abs/1812.07865} {arXiv:1812.07865 [gr-qc]} \BibitemShut
  {NoStop}%
\bibitem [{\citenamefont {Leung}\ \emph {et~al.}(2019)\citenamefont {Leung},
  \citenamefont {Nomoto},\ and\ \citenamefont {Blinnikov}}]{Leung:2019fgj}%
  \BibitemOpen
  \bibfield  {author} {\bibinfo {author} {\bibfnamefont {S.-C.}\ \bibnamefont
  {Leung}}, \bibinfo {author} {\bibfnamefont {K.}~\bibnamefont {Nomoto}}, \
  and\ \bibinfo {author} {\bibfnamefont {S.}~\bibnamefont {Blinnikov}},\ }\href
  {\doibase 10.3847/1538-4357/ab4fe5} {\bibfield  {journal} {\bibinfo
  {journal} {Astrophys. J.}\ }\textbf {\bibinfo {volume} {887}},\ \bibinfo
  {pages} {72} (\bibinfo {year} {2019})},\ \Eprint
  {http://arxiv.org/abs/1901.11136} {arXiv:1901.11136 [astro-ph.HE]}
  \BibitemShut {NoStop}%
\bibitem [{\citenamefont {Mapelli}\ \emph {et~al.}(2019)\citenamefont
  {Mapelli}, \citenamefont {Spera}, \citenamefont {Montanari}, \citenamefont
  {Limongi}, \citenamefont {Chieffi}, \citenamefont {Giacobbo}, \citenamefont
  {Bressan},\ and\ \citenamefont {Bouffanais}}]{Mapelli:2019ipt}%
  \BibitemOpen
  \bibfield  {author} {\bibinfo {author} {\bibfnamefont {M.}~\bibnamefont
  {Mapelli}}, \bibinfo {author} {\bibfnamefont {M.}~\bibnamefont {Spera}},
  \bibinfo {author} {\bibfnamefont {E.}~\bibnamefont {Montanari}}, \bibinfo
  {author} {\bibfnamefont {M.}~\bibnamefont {Limongi}}, \bibinfo {author}
  {\bibfnamefont {A.}~\bibnamefont {Chieffi}}, \bibinfo {author} {\bibfnamefont
  {N.}~\bibnamefont {Giacobbo}}, \bibinfo {author} {\bibfnamefont
  {A.}~\bibnamefont {Bressan}}, \ and\ \bibinfo {author} {\bibfnamefont
  {Y.}~\bibnamefont {Bouffanais}},\ }\href {\doibase 10.3847/1538-4357/ab584d}
  {\  (\bibinfo {year} {2019}),\ 10.3847/1538-4357/ab584d},\ \Eprint
  {http://arxiv.org/abs/1909.01371} {arXiv:1909.01371 [astro-ph.HE]}
  \BibitemShut {NoStop}%
\bibitem [{\citenamefont {Fryer}\ \emph {et~al.}(2012)\citenamefont {Fryer},
  \citenamefont {Belczynski}, \citenamefont {Wiktorowicz}, \citenamefont
  {Dominik}, \citenamefont {Kalogera},\ and\ \citenamefont
  {Holz}}]{Fryer:2011cx}%
  \BibitemOpen
  \bibfield  {author} {\bibinfo {author} {\bibfnamefont {C.~L.}\ \bibnamefont
  {Fryer}}, \bibinfo {author} {\bibfnamefont {K.}~\bibnamefont {Belczynski}},
  \bibinfo {author} {\bibfnamefont {G.}~\bibnamefont {Wiktorowicz}}, \bibinfo
  {author} {\bibfnamefont {M.}~\bibnamefont {Dominik}}, \bibinfo {author}
  {\bibfnamefont {V.}~\bibnamefont {Kalogera}}, \ and\ \bibinfo {author}
  {\bibfnamefont {D.~E.}\ \bibnamefont {Holz}},\ }\href {\doibase
  10.1088/0004-637X/749/1/91} {\bibfield  {journal} {\bibinfo  {journal}
  {Astrophys. J.}\ }\textbf {\bibinfo {volume} {749}},\ \bibinfo {pages} {91}
  (\bibinfo {year} {2012})},\ \Eprint {http://arxiv.org/abs/1110.1726}
  {arXiv:1110.1726 [astro-ph.SR]} \BibitemShut {NoStop}%
\bibitem [{\citenamefont {{Renzo}}\ \emph {et~al.}(2020)\citenamefont
  {{Renzo}}, \citenamefont {{Farmer}}, \citenamefont {{Justham}}, \citenamefont
  {{de Mink}}, \citenamefont {{G{\"o}tberg}},\ and\ \citenamefont
  {{Marchant}}}]{2020MNRAS.493.4333R}%
  \BibitemOpen
  \bibfield  {author} {\bibinfo {author} {\bibfnamefont {M.}~\bibnamefont
  {{Renzo}}}, \bibinfo {author} {\bibfnamefont {R.~J.}\ \bibnamefont
  {{Farmer}}}, \bibinfo {author} {\bibfnamefont {S.}~\bibnamefont {{Justham}}},
  \bibinfo {author} {\bibfnamefont {S.~E.}\ \bibnamefont {{de Mink}}}, \bibinfo
  {author} {\bibfnamefont {Y.}~\bibnamefont {{G{\"o}tberg}}}, \ and\ \bibinfo
  {author} {\bibfnamefont {P.}~\bibnamefont {{Marchant}}},\ }\href {\doibase
  10.1093/mnras/staa549} {\bibfield  {journal} {\bibinfo  {journal} {\mnras}\
  }\textbf {\bibinfo {volume} {493}},\ \bibinfo {pages} {4333} (\bibinfo {year}
  {2020})},\ \Eprint {http://arxiv.org/abs/2002.08200} {arXiv:2002.08200
  [astro-ph.SR]} \BibitemShut {NoStop}%
\bibitem [{\citenamefont {deBoer}\ \emph {et~al.}(2017)\citenamefont {deBoer}
  \emph {et~al.}}]{deBoer:2017ldl}%
  \BibitemOpen
  \bibfield  {author} {\bibinfo {author} {\bibfnamefont {R.}~\bibnamefont
  {deBoer}} \emph {et~al.},\ }\href {\doibase 10.1103/RevModPhys.89.035007}
  {\bibfield  {journal} {\bibinfo  {journal} {Rev. Mod. Phys.}\ }\textbf
  {\bibinfo {volume} {89}},\ \bibinfo {pages} {035007} (\bibinfo {year}
  {2017})},\ \Eprint {http://arxiv.org/abs/1709.03144} {arXiv:1709.03144
  [nucl-ex]} \BibitemShut {NoStop}%
\bibitem [{\citenamefont {Launey}\ \emph {et~al.}(2020)\citenamefont {Launey},
  \citenamefont {deBoer},\ and\ \citenamefont {Rocco}}]{nuclearconvo}%
  \BibitemOpen
  \bibfield  {author} {\bibinfo {author} {\bibfnamefont {K.}~\bibnamefont
  {Launey}}, \bibinfo {author} {\bibfnamefont {R.}~\bibnamefont {deBoer}}, \
  and\ \bibinfo {author} {\bibfnamefont {N.}~\bibnamefont {Rocco}},\
  }\href@noop {} {} (\bibinfo {year} {2020}),\ \bibinfo {note} {private
  conversation}\BibitemShut {NoStop}%
\bibitem [{\citenamefont {Roupas}\ and\ \citenamefont
  {Kazanas}(2019{\natexlab{a}})}]{Roupas:2018cvb}%
  \BibitemOpen
  \bibfield  {author} {\bibinfo {author} {\bibfnamefont {Z.}~\bibnamefont
  {Roupas}}\ and\ \bibinfo {author} {\bibfnamefont {D.}~\bibnamefont
  {Kazanas}},\ }\href {\doibase 10.1051/0004-6361/201834609} {\bibfield
  {journal} {\bibinfo  {journal} {Astron. Astrophys.}\ }\textbf {\bibinfo
  {volume} {621}},\ \bibinfo {pages} {L1} (\bibinfo {year}
  {2019}{\natexlab{a}})},\ \Eprint {http://arxiv.org/abs/1809.04126}
  {arXiv:1809.04126 [astro-ph.GA]} \BibitemShut {NoStop}%
\bibitem [{\citenamefont {Roupas}\ and\ \citenamefont
  {Kazanas}(2019{\natexlab{b}})}]{Roupas:2019dgx}%
  \BibitemOpen
  \bibfield  {author} {\bibinfo {author} {\bibfnamefont {Z.}~\bibnamefont
  {Roupas}}\ and\ \bibinfo {author} {\bibfnamefont {D.}~\bibnamefont
  {Kazanas}},\ }\href {\doibase 10.1051/0004-6361/201937002} {\bibfield
  {journal} {\bibinfo  {journal} {Astron. Astrophys.}\ }\textbf {\bibinfo
  {volume} {632}},\ \bibinfo {pages} {L8} (\bibinfo {year}
  {2019}{\natexlab{b}})},\ \Eprint {http://arxiv.org/abs/1911.03915}
  {arXiv:1911.03915} \BibitemShut {NoStop}%
\bibitem [{\citenamefont {{van Son}}\ \emph {et~al.}(2020)\citenamefont {{van
  Son}}, \citenamefont {{De Mink}}, \citenamefont {{Broekgaarden}},
  \citenamefont {{Renzo}}, \citenamefont {{Justham}}, \citenamefont
  {{Laplace}}, \citenamefont {{Mor{\'a}n-Fraile}}, \citenamefont {{Hendriks}},\
  and\ \citenamefont {{Farmer}}}]{2020ApJ...897..100V}%
  \BibitemOpen
  \bibfield  {author} {\bibinfo {author} {\bibfnamefont {L.~A.~C.}\
  \bibnamefont {{van Son}}}, \bibinfo {author} {\bibfnamefont {S.~E.}\
  \bibnamefont {{De Mink}}}, \bibinfo {author} {\bibfnamefont {F.~S.}\
  \bibnamefont {{Broekgaarden}}}, \bibinfo {author} {\bibfnamefont
  {M.}~\bibnamefont {{Renzo}}}, \bibinfo {author} {\bibfnamefont
  {S.}~\bibnamefont {{Justham}}}, \bibinfo {author} {\bibfnamefont
  {E.}~\bibnamefont {{Laplace}}}, \bibinfo {author} {\bibfnamefont
  {J.}~\bibnamefont {{Mor{\'a}n-Fraile}}}, \bibinfo {author} {\bibfnamefont
  {D.~D.}\ \bibnamefont {{Hendriks}}}, \ and\ \bibinfo {author} {\bibfnamefont
  {R.}~\bibnamefont {{Farmer}}},\ }\href {\doibase 10.3847/1538-4357/ab9809}
  {\bibfield  {journal} {\bibinfo  {journal} {\apj}\ }\textbf {\bibinfo
  {volume} {897}},\ \bibinfo {eid} {100} (\bibinfo {year} {2020})},\ \Eprint
  {http://arxiv.org/abs/2004.05187} {arXiv:2004.05187 [astro-ph.HE]}
  \BibitemShut {NoStop}%
\bibitem [{\citenamefont {Paxton}\ \emph {et~al.}(2011)\citenamefont {Paxton},
  \citenamefont {Bildsten}, \citenamefont {Dotter}, \citenamefont {Herwig},
  \citenamefont {Lesaffre},\ and\ \citenamefont {Timmes}}]{Paxton2011-MESA}%
  \BibitemOpen
  \bibfield  {author} {\bibinfo {author} {\bibfnamefont {B.}~\bibnamefont
  {Paxton}}, \bibinfo {author} {\bibfnamefont {L.}~\bibnamefont {Bildsten}},
  \bibinfo {author} {\bibfnamefont {A.}~\bibnamefont {Dotter}}, \bibinfo
  {author} {\bibfnamefont {F.}~\bibnamefont {Herwig}}, \bibinfo {author}
  {\bibfnamefont {P.}~\bibnamefont {Lesaffre}}, \ and\ \bibinfo {author}
  {\bibfnamefont {F.}~\bibnamefont {Timmes}},\ }\href {\doibase
  10.1088/0067-0049/192/1/3} {\bibfield  {journal} {\bibinfo  {journal}
  {Astrophys. J. Suppl.}\ }\textbf {\bibinfo {volume} {192}},\ \bibinfo {pages}
  {3} (\bibinfo {year} {2011})},\ \Eprint {http://arxiv.org/abs/1009.1622}
  {arXiv:1009.1622 [astro-ph.SR]} \BibitemShut {NoStop}%
\bibitem [{\citenamefont {Paxton}\ \emph {et~al.}(2015)\citenamefont {Paxton}
  \emph {et~al.}}]{Paxton:2015jva}%
  \BibitemOpen
  \bibfield  {author} {\bibinfo {author} {\bibfnamefont {B.}~\bibnamefont
  {Paxton}} \emph {et~al.},\ }\href {\doibase 10.1088/0067-0049/220/1/15}
  {\bibfield  {journal} {\bibinfo  {journal} {Astrophys. J. Suppl.}\ }\textbf
  {\bibinfo {volume} {220}},\ \bibinfo {pages} {15} (\bibinfo {year} {2015})},\
  \Eprint {http://arxiv.org/abs/1506.03146} {arXiv:1506.03146 [astro-ph.SR]}
  \BibitemShut {NoStop}%
\bibitem [{\citenamefont {Paxton}\ \emph {et~al.}(2018)\citenamefont {Paxton}
  \emph {et~al.}}]{Paxton2018-MESA}%
  \BibitemOpen
  \bibfield  {author} {\bibinfo {author} {\bibfnamefont {B.}~\bibnamefont
  {Paxton}} \emph {et~al.},\ }\href {\doibase 10.3847/1538-4365/aaa5a8}
  {\bibfield  {journal} {\bibinfo  {journal} {Astrophys. J. Suppl.}\ }\textbf
  {\bibinfo {volume} {234}},\ \bibinfo {pages} {34} (\bibinfo {year} {2018})},\
  \Eprint {http://arxiv.org/abs/1710.08424} {arXiv:1710.08424 [astro-ph.SR]}
  \BibitemShut {NoStop}%
\bibitem [{\citenamefont {Aprile}\ \emph {et~al.}(2020)\citenamefont {Aprile}
  \emph {et~al.}}]{Aprile:2020tmw}%
  \BibitemOpen
  \bibfield  {author} {\bibinfo {author} {\bibfnamefont {E.}~\bibnamefont
  {Aprile}} \emph {et~al.} (\bibinfo {collaboration} {XENON}),\ }\href@noop {}
  {\  (\bibinfo {year} {2020})},\ \Eprint {http://arxiv.org/abs/2006.09721}
  {arXiv:2006.09721 [hep-ex]} \BibitemShut {NoStop}%
\bibitem [{\citenamefont {Di~Luzio}\ \emph {et~al.}(2020)\citenamefont
  {Di~Luzio}, \citenamefont {Fedele}, \citenamefont {Giannotti}, \citenamefont
  {Mescia},\ and\ \citenamefont {Nardi}}]{DiLuzio:2020jjp}%
  \BibitemOpen
  \bibfield  {author} {\bibinfo {author} {\bibfnamefont {L.}~\bibnamefont
  {Di~Luzio}}, \bibinfo {author} {\bibfnamefont {M.}~\bibnamefont {Fedele}},
  \bibinfo {author} {\bibfnamefont {M.}~\bibnamefont {Giannotti}}, \bibinfo
  {author} {\bibfnamefont {F.}~\bibnamefont {Mescia}}, \ and\ \bibinfo {author}
  {\bibfnamefont {E.}~\bibnamefont {Nardi}},\ }\href@noop {} {\  (\bibinfo
  {year} {2020})},\ \Eprint {http://arxiv.org/abs/2006.12487} {arXiv:2006.12487
  [hep-ph]} \BibitemShut {NoStop}%
\bibitem [{\citenamefont {Anastassopoulos}\ \emph {et~al.}(2017)\citenamefont
  {Anastassopoulos} \emph {et~al.}}]{Anastassopoulos:2017ftl}%
  \BibitemOpen
  \bibfield  {author} {\bibinfo {author} {\bibfnamefont {V.}~\bibnamefont
  {Anastassopoulos}} \emph {et~al.} (\bibinfo {collaboration} {CAST}),\ }\href
  {\doibase 10.1038/nphys4109} {\bibfield  {journal} {\bibinfo  {journal}
  {Nature Phys.}\ }\textbf {\bibinfo {volume} {13}},\ \bibinfo {pages} {584}
  (\bibinfo {year} {2017})},\ \Eprint {http://arxiv.org/abs/1705.02290}
  {arXiv:1705.02290 [hep-ex]} \BibitemShut {NoStop}%
\bibitem [{\citenamefont {{Vinyoles}}\ \emph {et~al.}(2015)\citenamefont
  {{Vinyoles}}, \citenamefont {{Serenelli}}, \citenamefont {{Villante}},
  \citenamefont {{Basu}}, \citenamefont {{Redondo}},\ and\ \citenamefont
  {{Isern}}}]{2015JCAP...10..015V}%
  \BibitemOpen
  \bibfield  {author} {\bibinfo {author} {\bibfnamefont {N.}~\bibnamefont
  {{Vinyoles}}}, \bibinfo {author} {\bibfnamefont {A.}~\bibnamefont
  {{Serenelli}}}, \bibinfo {author} {\bibfnamefont {F.~L.}\ \bibnamefont
  {{Villante}}}, \bibinfo {author} {\bibfnamefont {S.}~\bibnamefont {{Basu}}},
  \bibinfo {author} {\bibfnamefont {J.}~\bibnamefont {{Redondo}}}, \ and\
  \bibinfo {author} {\bibfnamefont {J.}~\bibnamefont {{Isern}}},\ }\href
  {\doibase 10.1088/1475-7516/2015/10/015} {\bibfield  {journal} {\bibinfo
  {journal} {\jcap}\ }\textbf {\bibinfo {volume} {2015}},\ \bibinfo {eid} {015}
  (\bibinfo {year} {2015})},\ \Eprint {http://arxiv.org/abs/1501.01639}
  {arXiv:1501.01639 [astro-ph.SR]} \BibitemShut {NoStop}%
\bibitem [{\citenamefont {An}\ \emph {et~al.}(2013)\citenamefont {An},
  \citenamefont {Pospelov},\ and\ \citenamefont {Pradler}}]{An:2013yfc}%
  \BibitemOpen
  \bibfield  {author} {\bibinfo {author} {\bibfnamefont {H.}~\bibnamefont
  {An}}, \bibinfo {author} {\bibfnamefont {M.}~\bibnamefont {Pospelov}}, \ and\
  \bibinfo {author} {\bibfnamefont {J.}~\bibnamefont {Pradler}},\ }\href
  {\doibase 10.1016/j.physletb.2013.07.008} {\bibfield  {journal} {\bibinfo
  {journal} {Phys. Lett. B}\ }\textbf {\bibinfo {volume} {725}},\ \bibinfo
  {pages} {190} (\bibinfo {year} {2013})},\ \Eprint
  {http://arxiv.org/abs/1302.3884} {arXiv:1302.3884 [hep-ph]} \BibitemShut
  {NoStop}%
\bibitem [{\citenamefont {Redondo}\ and\ \citenamefont
  {Raffelt}(2013)}]{Redondo:2013lna}%
  \BibitemOpen
  \bibfield  {author} {\bibinfo {author} {\bibfnamefont {J.}~\bibnamefont
  {Redondo}}\ and\ \bibinfo {author} {\bibfnamefont {G.}~\bibnamefont
  {Raffelt}},\ }\href {\doibase 10.1088/1475-7516/2013/08/034} {\bibfield
  {journal} {\bibinfo  {journal} {JCAP}\ }\textbf {\bibinfo {volume} {08}},\
  \bibinfo {pages} {034} (\bibinfo {year} {2013})},\ \Eprint
  {http://arxiv.org/abs/1305.2920} {arXiv:1305.2920 [hep-ph]} \BibitemShut
  {NoStop}%
\bibitem [{\citenamefont {Gao}\ \emph {et~al.}(2020)\citenamefont {Gao},
  \citenamefont {Liu}, \citenamefont {Wang}, \citenamefont {Wang},
  \citenamefont {Xue},\ and\ \citenamefont {Zhong}}]{Gao:2020wer}%
  \BibitemOpen
  \bibfield  {author} {\bibinfo {author} {\bibfnamefont {C.}~\bibnamefont
  {Gao}}, \bibinfo {author} {\bibfnamefont {J.}~\bibnamefont {Liu}}, \bibinfo
  {author} {\bibfnamefont {L.-T.}\ \bibnamefont {Wang}}, \bibinfo {author}
  {\bibfnamefont {X.-P.}\ \bibnamefont {Wang}}, \bibinfo {author}
  {\bibfnamefont {W.}~\bibnamefont {Xue}}, \ and\ \bibinfo {author}
  {\bibfnamefont {Y.-M.}\ \bibnamefont {Zhong}},\ }\href@noop {} {\  (\bibinfo
  {year} {2020})},\ \Eprint {http://arxiv.org/abs/2006.14598} {arXiv:2006.14598
  [hep-ph]} \BibitemShut {NoStop}%
\bibitem [{\citenamefont {Copeland}\ \emph {et~al.}(2006)\citenamefont
  {Copeland}, \citenamefont {Sami},\ and\ \citenamefont
  {Tsujikawa}}]{Copeland:2006wr}%
  \BibitemOpen
  \bibfield  {author} {\bibinfo {author} {\bibfnamefont {E.~J.}\ \bibnamefont
  {Copeland}}, \bibinfo {author} {\bibfnamefont {M.}~\bibnamefont {Sami}}, \
  and\ \bibinfo {author} {\bibfnamefont {S.}~\bibnamefont {Tsujikawa}},\ }\href
  {\doibase 10.1142/S021827180600942X} {\bibfield  {journal} {\bibinfo
  {journal} {Int. J. Mod. Phys. D}\ }\textbf {\bibinfo {volume} {15}},\
  \bibinfo {pages} {1753} (\bibinfo {year} {2006})},\ \Eprint
  {http://arxiv.org/abs/hep-th/0603057} {arXiv:hep-th/0603057} \BibitemShut
  {NoStop}%
\bibitem [{\citenamefont {Burgess}(2015)}]{Burgess:2013ara}%
  \BibitemOpen
  \bibfield  {author} {\bibinfo {author} {\bibfnamefont {C.}~\bibnamefont
  {Burgess}},\ }in\ \href {\doibase 10.1093/acprof:oso/9780198728856.003.0004}
  {\emph {\bibinfo {booktitle} {{100e Ecole d'Ete de Physique: Post-Planck
  Cosmology}}}}\ (\bibinfo {year} {2015})\ pp.\ \bibinfo {pages} {149--197},\
  \Eprint {http://arxiv.org/abs/1309.4133} {arXiv:1309.4133 [hep-th]}
  \BibitemShut {NoStop}%
\bibitem [{\citenamefont {Padilla}(2015)}]{Padilla:2015aaa}%
  \BibitemOpen
  \bibfield  {author} {\bibinfo {author} {\bibfnamefont {A.}~\bibnamefont
  {Padilla}},\ }\href@noop {} {\  (\bibinfo {year} {2015})},\ \Eprint
  {http://arxiv.org/abs/1502.05296} {arXiv:1502.05296 [hep-th]} \BibitemShut
  {NoStop}%
\bibitem [{\citenamefont {Khoury}\ \emph {et~al.}(2018)\citenamefont {Khoury},
  \citenamefont {Sakstein},\ and\ \citenamefont {Solomon}}]{Khoury:2018vdv}%
  \BibitemOpen
  \bibfield  {author} {\bibinfo {author} {\bibfnamefont {J.}~\bibnamefont
  {Khoury}}, \bibinfo {author} {\bibfnamefont {J.}~\bibnamefont {Sakstein}}, \
  and\ \bibinfo {author} {\bibfnamefont {A.~R.}\ \bibnamefont {Solomon}},\
  }\href {\doibase 10.1088/1475-7516/2018/08/024} {\bibfield  {journal}
  {\bibinfo  {journal} {JCAP}\ }\textbf {\bibinfo {volume} {08}},\ \bibinfo
  {pages} {024} (\bibinfo {year} {2018})},\ \Eprint
  {http://arxiv.org/abs/1805.05937} {arXiv:1805.05937 [hep-th]} \BibitemShut
  {NoStop}%
\bibitem [{\citenamefont {Mori}\ \emph {et~al.}(2020)\citenamefont {Mori},
  \citenamefont {Balantekin}, \citenamefont {Kajino},\ and\ \citenamefont
  {Famiano}}]{Mori:2020niw}%
  \BibitemOpen
  \bibfield  {author} {\bibinfo {author} {\bibfnamefont {K.}~\bibnamefont
  {Mori}}, \bibinfo {author} {\bibfnamefont {A.~B.}\ \bibnamefont
  {Balantekin}}, \bibinfo {author} {\bibfnamefont {T.}~\bibnamefont {Kajino}},
  \ and\ \bibinfo {author} {\bibfnamefont {M.~A.}\ \bibnamefont {Famiano}},\
  }\href@noop {} {\  (\bibinfo {year} {2020})},\ \Eprint
  {http://arxiv.org/abs/2008.08393} {arXiv:2008.08393 [astro-ph.SR]}
  \BibitemShut {NoStop}%
\bibitem [{\citenamefont {Tan}\ \emph {et~al.}(2020)\citenamefont {Tan} \emph
  {et~al.}}]{Tan:2020vpf}%
  \BibitemOpen
  \bibfield  {author} {\bibinfo {author} {\bibfnamefont {W.-H.}\ \bibnamefont
  {Tan}} \emph {et~al.},\ }\href {\doibase 10.1103/PhysRevLett.124.051301}
  {\bibfield  {journal} {\bibinfo  {journal} {Phys. Rev. Lett.}\ }\textbf
  {\bibinfo {volume} {124}},\ \bibinfo {pages} {051301} (\bibinfo {year}
  {2020})}\BibitemShut {NoStop}%
\bibitem [{\citenamefont {Lee}\ \emph {et~al.}(2020)\citenamefont {Lee},
  \citenamefont {Adelberger}, \citenamefont {Cook}, \citenamefont {Fleischer},\
  and\ \citenamefont {Heckel}}]{Lee:2020zjt}%
  \BibitemOpen
  \bibfield  {author} {\bibinfo {author} {\bibfnamefont {J.}~\bibnamefont
  {Lee}}, \bibinfo {author} {\bibfnamefont {E.}~\bibnamefont {Adelberger}},
  \bibinfo {author} {\bibfnamefont {T.}~\bibnamefont {Cook}}, \bibinfo {author}
  {\bibfnamefont {S.}~\bibnamefont {Fleischer}}, \ and\ \bibinfo {author}
  {\bibfnamefont {B.}~\bibnamefont {Heckel}},\ }\href {\doibase
  10.1103/PhysRevLett.124.101101} {\bibfield  {journal} {\bibinfo  {journal}
  {Phys. Rev. Lett.}\ }\textbf {\bibinfo {volume} {124}},\ \bibinfo {pages}
  {101101} (\bibinfo {year} {2020})},\ \Eprint
  {http://arxiv.org/abs/2002.11761} {arXiv:2002.11761 [hep-ex]} \BibitemShut
  {NoStop}%
\bibitem [{\citenamefont {Sirunyan}\ \emph {et~al.}(2018)\citenamefont
  {Sirunyan} \emph {et~al.}}]{Sirunyan:2017jix}%
  \BibitemOpen
  \bibfield  {author} {\bibinfo {author} {\bibfnamefont {A.~M.}\ \bibnamefont
  {Sirunyan}} \emph {et~al.} (\bibinfo {collaboration} {CMS}),\ }\href
  {\doibase 10.1103/PhysRevD.97.092005} {\bibfield  {journal} {\bibinfo
  {journal} {Phys. Rev. D}\ }\textbf {\bibinfo {volume} {97}},\ \bibinfo
  {pages} {092005} (\bibinfo {year} {2018})},\ \Eprint
  {http://arxiv.org/abs/1712.02345} {arXiv:1712.02345 [hep-ex]} \BibitemShut
  {NoStop}%
\bibitem [{\citenamefont {Balantekin}\ and\ \citenamefont
  {Vassh}(2014)}]{Balantekin:2013sda}%
  \BibitemOpen
  \bibfield  {author} {\bibinfo {author} {\bibfnamefont {A.}~\bibnamefont
  {Balantekin}}\ and\ \bibinfo {author} {\bibfnamefont {N.}~\bibnamefont
  {Vassh}},\ }\href {\doibase 10.1103/PhysRevD.89.073013} {\bibfield  {journal}
  {\bibinfo  {journal} {Phys. Rev. D}\ }\textbf {\bibinfo {volume} {89}},\
  \bibinfo {pages} {073013} (\bibinfo {year} {2014})},\ \Eprint
  {http://arxiv.org/abs/1312.6858} {arXiv:1312.6858 [hep-ph]} \BibitemShut
  {NoStop}%
\bibitem [{\citenamefont {Barger}\ \emph {et~al.}(1999)\citenamefont {Barger},
  \citenamefont {Han}, \citenamefont {Kao},\ and\ \citenamefont
  {Zhang}}]{Barger:1999jf}%
  \BibitemOpen
  \bibfield  {author} {\bibinfo {author} {\bibfnamefont {V.~D.}\ \bibnamefont
  {Barger}}, \bibinfo {author} {\bibfnamefont {T.}~\bibnamefont {Han}},
  \bibinfo {author} {\bibfnamefont {C.}~\bibnamefont {Kao}}, \ and\ \bibinfo
  {author} {\bibfnamefont {R.-J.}\ \bibnamefont {Zhang}},\ }\href {\doibase
  10.1016/S0370-2693(99)00795-9} {\bibfield  {journal} {\bibinfo  {journal}
  {Phys. Lett. B}\ }\textbf {\bibinfo {volume} {461}},\ \bibinfo {pages} {34}
  (\bibinfo {year} {1999})},\ \Eprint {http://arxiv.org/abs/hep-ph/9905474}
  {arXiv:hep-ph/9905474} \BibitemShut {NoStop}%
\bibitem [{\citenamefont {Heger}\ \emph {et~al.}(2009)\citenamefont {Heger},
  \citenamefont {Friedland}, \citenamefont {Giannotti},\ and\ \citenamefont
  {Cirigliano}}]{Heger:2008er}%
  \BibitemOpen
  \bibfield  {author} {\bibinfo {author} {\bibfnamefont {A.}~\bibnamefont
  {Heger}}, \bibinfo {author} {\bibfnamefont {A.}~\bibnamefont {Friedland}},
  \bibinfo {author} {\bibfnamefont {M.}~\bibnamefont {Giannotti}}, \ and\
  \bibinfo {author} {\bibfnamefont {V.}~\bibnamefont {Cirigliano}},\ }\href
  {\doibase 10.1088/0004-637X/696/1/608} {\bibfield  {journal} {\bibinfo
  {journal} {Astrophys. J.}\ }\textbf {\bibinfo {volume} {696}},\ \bibinfo
  {pages} {608} (\bibinfo {year} {2009})},\ \Eprint
  {http://arxiv.org/abs/0809.4703} {arXiv:0809.4703 [astro-ph]} \BibitemShut
  {NoStop}%
\bibitem [{\citenamefont {Raffelt}(1999)}]{Raffelt:1999tx}%
  \BibitemOpen
  \bibfield  {author} {\bibinfo {author} {\bibfnamefont {G.~G.}\ \bibnamefont
  {Raffelt}},\ }\href {\doibase 10.1146/annurev.nucl.49.1.163} {\bibfield
  {journal} {\bibinfo  {journal} {Ann. Rev. Nucl. Part. Sci.}\ }\textbf
  {\bibinfo {volume} {49}},\ \bibinfo {pages} {163} (\bibinfo {year} {1999})},\
  \Eprint {http://arxiv.org/abs/hep-ph/9903472} {arXiv:hep-ph/9903472}
  \BibitemShut {NoStop}%
\bibitem [{\citenamefont {Giunti}\ and\ \citenamefont
  {Studenikin}(2015)}]{Giunti:2014ixa}%
  \BibitemOpen
  \bibfield  {author} {\bibinfo {author} {\bibfnamefont {C.}~\bibnamefont
  {Giunti}}\ and\ \bibinfo {author} {\bibfnamefont {A.}~\bibnamefont
  {Studenikin}},\ }\href {\doibase 10.1103/RevModPhys.87.531} {\bibfield
  {journal} {\bibinfo  {journal} {Rev. Mod. Phys.}\ }\textbf {\bibinfo {volume}
  {87}},\ \bibinfo {pages} {531} (\bibinfo {year} {2015})},\ \Eprint
  {http://arxiv.org/abs/1403.6344} {arXiv:1403.6344 [hep-ph]} \BibitemShut
  {NoStop}%
\bibitem [{\citenamefont {Joyce}\ \emph {et~al.}(2015)\citenamefont {Joyce},
  \citenamefont {Jain}, \citenamefont {Khoury},\ and\ \citenamefont
  {Trodden}}]{Joyce:2014kja}%
  \BibitemOpen
  \bibfield  {author} {\bibinfo {author} {\bibfnamefont {A.}~\bibnamefont
  {Joyce}}, \bibinfo {author} {\bibfnamefont {B.}~\bibnamefont {Jain}},
  \bibinfo {author} {\bibfnamefont {J.}~\bibnamefont {Khoury}}, \ and\ \bibinfo
  {author} {\bibfnamefont {M.}~\bibnamefont {Trodden}},\ }\href {\doibase
  10.1016/j.physrep.2014.12.002} {\bibfield  {journal} {\bibinfo  {journal}
  {Phys. Rept.}\ }\textbf {\bibinfo {volume} {568}},\ \bibinfo {pages} {1}
  (\bibinfo {year} {2015})},\ \Eprint {http://arxiv.org/abs/1407.0059}
  {arXiv:1407.0059 [astro-ph.CO]} \BibitemShut {NoStop}%
\bibitem [{\citenamefont {Burrage}\ and\ \citenamefont
  {Sakstein}(2018)}]{Burrage:2017qrf}%
  \BibitemOpen
  \bibfield  {author} {\bibinfo {author} {\bibfnamefont {C.}~\bibnamefont
  {Burrage}}\ and\ \bibinfo {author} {\bibfnamefont {J.}~\bibnamefont
  {Sakstein}},\ }\href {\doibase 10.1007/s41114-018-0011-x} {\bibfield
  {journal} {\bibinfo  {journal} {Living Rev. Rel.}\ }\textbf {\bibinfo
  {volume} {21}},\ \bibinfo {pages} {1} (\bibinfo {year} {2018})},\ \Eprint
  {http://arxiv.org/abs/1709.09071} {arXiv:1709.09071 [astro-ph.CO]}
  \BibitemShut {NoStop}%
\bibitem [{\citenamefont {Sakstein}(2018)}]{Sakstein2018-testScreened}%
  \BibitemOpen
  \bibfield  {author} {\bibinfo {author} {\bibfnamefont {J.}~\bibnamefont
  {Sakstein}},\ }\href {\doibase 10.1142/S0218271818480085} {\bibfield
  {journal} {\bibinfo  {journal} {Int. J. Mod. Phys. D}\ }\textbf {\bibinfo
  {volume} {27}},\ \bibinfo {pages} {1848008} (\bibinfo {year} {2018})},\
  \Eprint {http://arxiv.org/abs/2002.04194} {arXiv:2002.04194 [astro-ph.CO]}
  \BibitemShut {NoStop}%
\bibitem [{\citenamefont {Baker}\ \emph {et~al.}(2019)\citenamefont {Baker}
  \emph {et~al.}}]{Baker:2019gxo}%
  \BibitemOpen
  \bibfield  {author} {\bibinfo {author} {\bibfnamefont {T.}~\bibnamefont
  {Baker}} \emph {et~al.},\ }\href@noop {} {\  (\bibinfo {year} {2019})},\
  \Eprint {http://arxiv.org/abs/1908.03430} {arXiv:1908.03430 [astro-ph.CO]}
  \BibitemShut {NoStop}%
\bibitem [{\citenamefont {Sakstein}\ \emph {et~al.}(2019)\citenamefont
  {Sakstein}, \citenamefont {Desmond},\ and\ \citenamefont
  {Jain}}]{Sakstein2019-Screened}%
  \BibitemOpen
  \bibfield  {author} {\bibinfo {author} {\bibfnamefont {J.}~\bibnamefont
  {Sakstein}}, \bibinfo {author} {\bibfnamefont {H.}~\bibnamefont {Desmond}}, \
  and\ \bibinfo {author} {\bibfnamefont {B.}~\bibnamefont {Jain}},\ }\href
  {\doibase 10.1103/PhysRevD.100.104035} {\bibfield  {journal} {\bibinfo
  {journal} {Phys. Rev. D}\ }\textbf {\bibinfo {volume} {100}},\ \bibinfo
  {pages} {104035} (\bibinfo {year} {2019})},\ \Eprint
  {http://arxiv.org/abs/1907.03775} {arXiv:1907.03775 [astro-ph.CO]}
  \BibitemShut {NoStop}%
\bibitem [{\citenamefont {Desmond}\ \emph {et~al.}(2019)\citenamefont
  {Desmond}, \citenamefont {Jain},\ and\ \citenamefont
  {Sakstein}}]{Desmond2019-resolveHT}%
  \BibitemOpen
  \bibfield  {author} {\bibinfo {author} {\bibfnamefont {H.}~\bibnamefont
  {Desmond}}, \bibinfo {author} {\bibfnamefont {B.}~\bibnamefont {Jain}}, \
  and\ \bibinfo {author} {\bibfnamefont {J.}~\bibnamefont {Sakstein}},\ }\href
  {\doibase 10.1103/PhysRevD.100.043537} {\bibfield  {journal} {\bibinfo
  {journal} {Phys. Rev. D}\ }\textbf {\bibinfo {volume} {100}},\ \bibinfo
  {pages} {043537} (\bibinfo {year} {2019})},\ \bibinfo {note} {[Erratum:
  Phys.Rev.D 101, 069904 (2020), Erratum: Phys.Rev.D 101, 129901 (2020)]},\
  \Eprint {http://arxiv.org/abs/1907.03778} {arXiv:1907.03778 [astro-ph.CO]}
  \BibitemShut {NoStop}%
\bibitem [{\citenamefont {Desmond}\ and\ \citenamefont
  {Sakstein}(2020)}]{Desmond2020-Screened}%
  \BibitemOpen
  \bibfield  {author} {\bibinfo {author} {\bibfnamefont {H.}~\bibnamefont
  {Desmond}}\ and\ \bibinfo {author} {\bibfnamefont {J.}~\bibnamefont
  {Sakstein}},\ }\href {\doibase 10.1103/PhysRevD.102.023007} {\bibfield
  {journal} {\bibinfo  {journal} {Phys. Rev. D}\ }\textbf {\bibinfo {volume}
  {102}},\ \bibinfo {pages} {023007} (\bibinfo {year} {2020})},\ \Eprint
  {http://arxiv.org/abs/2003.12876} {arXiv:2003.12876 [astro-ph.CO]}
  \BibitemShut {NoStop}%
\bibitem [{\citenamefont {Hui}\ and\ \citenamefont
  {Nicolis}(2013)}]{Hui:2012qt}%
  \BibitemOpen
  \bibfield  {author} {\bibinfo {author} {\bibfnamefont {L.}~\bibnamefont
  {Hui}}\ and\ \bibinfo {author} {\bibfnamefont {A.}~\bibnamefont {Nicolis}},\
  }\href {\doibase 10.1103/PhysRevLett.110.241104} {\bibfield  {journal}
  {\bibinfo  {journal} {Phys. Rev. Lett.}\ }\textbf {\bibinfo {volume} {110}},\
  \bibinfo {pages} {241104} (\bibinfo {year} {2013})},\ \Eprint
  {http://arxiv.org/abs/1202.1296} {arXiv:1202.1296 [hep-th]} \BibitemShut
  {NoStop}%
\bibitem [{\citenamefont {Sotiriou}\ and\ \citenamefont
  {Zhou}(2014)}]{Sotiriou:2013qea}%
  \BibitemOpen
  \bibfield  {author} {\bibinfo {author} {\bibfnamefont {T.~P.}\ \bibnamefont
  {Sotiriou}}\ and\ \bibinfo {author} {\bibfnamefont {S.-Y.}\ \bibnamefont
  {Zhou}},\ }\href {\doibase 10.1103/PhysRevLett.112.251102} {\bibfield
  {journal} {\bibinfo  {journal} {Phys. Rev. Lett.}\ }\textbf {\bibinfo
  {volume} {112}},\ \bibinfo {pages} {251102} (\bibinfo {year} {2014})},\
  \Eprint {http://arxiv.org/abs/1312.3622} {arXiv:1312.3622 [gr-qc]}
  \BibitemShut {NoStop}%
\bibitem [{\citenamefont {Koyama}\ and\ \citenamefont
  {Sakstein}(2015)}]{Koyama:2015oma}%
  \BibitemOpen
  \bibfield  {author} {\bibinfo {author} {\bibfnamefont {K.}~\bibnamefont
  {Koyama}}\ and\ \bibinfo {author} {\bibfnamefont {J.}~\bibnamefont
  {Sakstein}},\ }\href {\doibase 10.1103/PhysRevD.91.124066} {\bibfield
  {journal} {\bibinfo  {journal} {Phys. Rev. D}\ }\textbf {\bibinfo {volume}
  {91}},\ \bibinfo {pages} {124066} (\bibinfo {year} {2015})},\ \Eprint
  {http://arxiv.org/abs/1502.06872} {arXiv:1502.06872 [astro-ph.CO]}
  \BibitemShut {NoStop}%
\bibitem [{\citenamefont {Saito}\ \emph {et~al.}(2015)\citenamefont {Saito},
  \citenamefont {Yamauchi}, \citenamefont {Mizuno}, \citenamefont {Gleyzes},\
  and\ \citenamefont {Langlois}}]{Saito:2015fza}%
  \BibitemOpen
  \bibfield  {author} {\bibinfo {author} {\bibfnamefont {R.}~\bibnamefont
  {Saito}}, \bibinfo {author} {\bibfnamefont {D.}~\bibnamefont {Yamauchi}},
  \bibinfo {author} {\bibfnamefont {S.}~\bibnamefont {Mizuno}}, \bibinfo
  {author} {\bibfnamefont {J.}~\bibnamefont {Gleyzes}}, \ and\ \bibinfo
  {author} {\bibfnamefont {D.}~\bibnamefont {Langlois}},\ }\href {\doibase
  10.1088/1475-7516/2015/06/008} {\bibfield  {journal} {\bibinfo  {journal}
  {JCAP}\ }\textbf {\bibinfo {volume} {1506}},\ \bibinfo {pages} {008}
  (\bibinfo {year} {2015})},\ \Eprint {http://arxiv.org/abs/1503.01448}
  {arXiv:1503.01448 [gr-qc]} \BibitemShut {NoStop}%
\bibitem [{\citenamefont {Sakstein}(2015{\natexlab{a}})}]{Sakstein:2015zoa}%
  \BibitemOpen
  \bibfield  {author} {\bibinfo {author} {\bibfnamefont {J.}~\bibnamefont
  {Sakstein}},\ }\href {\doibase 10.1103/PhysRevLett.115.201101} {\bibfield
  {journal} {\bibinfo  {journal} {Phys. Rev. Lett.}\ }\textbf {\bibinfo
  {volume} {115}},\ \bibinfo {pages} {201101} (\bibinfo {year}
  {2015}{\natexlab{a}})},\ \Eprint {http://arxiv.org/abs/1510.05964}
  {arXiv:1510.05964 [astro-ph.CO]} \BibitemShut {NoStop}%
\bibitem [{\citenamefont {Sakstein}(2015{\natexlab{b}})}]{Sakstein:2015aac}%
  \BibitemOpen
  \bibfield  {author} {\bibinfo {author} {\bibfnamefont {J.}~\bibnamefont
  {Sakstein}},\ }\href {\doibase 10.1103/PhysRevD.92.124045} {\bibfield
  {journal} {\bibinfo  {journal} {Phys. Rev. D}\ }\textbf {\bibinfo {volume}
  {92}},\ \bibinfo {pages} {124045} (\bibinfo {year} {2015}{\natexlab{b}})},\
  \Eprint {http://arxiv.org/abs/1511.01685} {arXiv:1511.01685 [astro-ph.CO]}
  \BibitemShut {NoStop}%
\bibitem [{\citenamefont {Sakstein}\ \emph
  {et~al.}(2017{\natexlab{a}})\citenamefont {Sakstein}, \citenamefont
  {Kenna-Allison},\ and\ \citenamefont {Koyama}}]{Sakstein:2016lyj}%
  \BibitemOpen
  \bibfield  {author} {\bibinfo {author} {\bibfnamefont {J.}~\bibnamefont
  {Sakstein}}, \bibinfo {author} {\bibfnamefont {M.}~\bibnamefont
  {Kenna-Allison}}, \ and\ \bibinfo {author} {\bibfnamefont {K.}~\bibnamefont
  {Koyama}},\ }\href {\doibase 10.1088/1475-7516/2017/03/007} {\bibfield
  {journal} {\bibinfo  {journal} {JCAP}\ }\textbf {\bibinfo {volume} {03}},\
  \bibinfo {pages} {007} (\bibinfo {year} {2017}{\natexlab{a}})},\ \Eprint
  {http://arxiv.org/abs/1611.01062} {arXiv:1611.01062 [gr-qc]} \BibitemShut
  {NoStop}%
\bibitem [{\citenamefont {Babichev}\ \emph {et~al.}(2016)\citenamefont
  {Babichev}, \citenamefont {Koyama}, \citenamefont {Langlois}, \citenamefont
  {Saito},\ and\ \citenamefont {Sakstein}}]{Babichev:2016jom}%
  \BibitemOpen
  \bibfield  {author} {\bibinfo {author} {\bibfnamefont {E.}~\bibnamefont
  {Babichev}}, \bibinfo {author} {\bibfnamefont {K.}~\bibnamefont {Koyama}},
  \bibinfo {author} {\bibfnamefont {D.}~\bibnamefont {Langlois}}, \bibinfo
  {author} {\bibfnamefont {R.}~\bibnamefont {Saito}}, \ and\ \bibinfo {author}
  {\bibfnamefont {J.}~\bibnamefont {Sakstein}},\ }\href {\doibase
  10.1088/0264-9381/33/23/235014} {\bibfield  {journal} {\bibinfo  {journal}
  {Class. Quant. Grav.}\ }\textbf {\bibinfo {volume} {33}},\ \bibinfo {pages}
  {235014} (\bibinfo {year} {2016})},\ \Eprint
  {http://arxiv.org/abs/1606.06627} {arXiv:1606.06627 [gr-qc]} \BibitemShut
  {NoStop}%
\bibitem [{\citenamefont {Sakstein}\ \emph
  {et~al.}(2017{\natexlab{b}})\citenamefont {Sakstein}, \citenamefont
  {Babichev}, \citenamefont {Koyama}, \citenamefont {Langlois},\ and\
  \citenamefont {Saito}}]{Sakstein:2016oel}%
  \BibitemOpen
  \bibfield  {author} {\bibinfo {author} {\bibfnamefont {J.}~\bibnamefont
  {Sakstein}}, \bibinfo {author} {\bibfnamefont {E.}~\bibnamefont {Babichev}},
  \bibinfo {author} {\bibfnamefont {K.}~\bibnamefont {Koyama}}, \bibinfo
  {author} {\bibfnamefont {D.}~\bibnamefont {Langlois}}, \ and\ \bibinfo
  {author} {\bibfnamefont {R.}~\bibnamefont {Saito}},\ }\href {\doibase
  10.1103/PhysRevD.95.064013} {\bibfield  {journal} {\bibinfo  {journal} {Phys.
  Rev.}\ }\textbf {\bibinfo {volume} {D95}},\ \bibinfo {pages} {064013}
  (\bibinfo {year} {2017}{\natexlab{b}})},\ \Eprint
  {http://arxiv.org/abs/1612.04263} {arXiv:1612.04263 [gr-qc]} \BibitemShut
  {NoStop}%
\bibitem [{\citenamefont {Saltas}\ and\ \citenamefont
  {Lopes}(2019)}]{Saltas:2019ius}%
  \BibitemOpen
  \bibfield  {author} {\bibinfo {author} {\bibfnamefont {I.~D.}\ \bibnamefont
  {Saltas}}\ and\ \bibinfo {author} {\bibfnamefont {I.}~\bibnamefont {Lopes}},\
  }\href {\doibase 10.1103/PhysRevLett.123.091103} {\bibfield  {journal}
  {\bibinfo  {journal} {Phys. Rev. Lett.}\ }\textbf {\bibinfo {volume} {123}},\
  \bibinfo {pages} {091103} (\bibinfo {year} {2019})},\ \Eprint
  {http://arxiv.org/abs/1909.02552} {arXiv:1909.02552 [astro-ph.CO]}
  \BibitemShut {NoStop}%
\end{thebibliography}%
\end{document}